\documentclass[12pt]{article}
\usepackage{amssymb}
\begin{document}
\renewcommand{\Bbb}{\mathbb}

%\batchmode

\thispagestyle{empty}

%\input{psfig}

%%%%%%%%%%%%%%%%%%%%%%%%%%%%%%%%%%%%%%%%%%%%%%%%%%%%%%%%%%%%%%%%%%%%%%%%%
%                            GREEK                                      %
%%%%%%%%%%%%%%%%%%%%%%%%%%%%%%%%%%%%%%%%%%%%%%%%%%%%%%%%%%%%%%%%%%%%%%%%%
\newcommand{\al}{\alpha}
\newcommand{\bet}{\beta}
\newcommand{\ga}{\gamma}
\newcommand{\del}{\delta}
\newcommand{\ep}{\epsilon}
\newcommand{\epx}{\varepsilon}
\newcommand{\ze}{\zeta}
\renewcommand{\th}{\theta}
\newcommand{\thx}{\vartheta}
\newcommand{\io}{\iota}
\newcommand{\la}{\lambda}
\newcommand{\ka}{\kappa}
\newcommand{\pix}{\varpi}
\newcommand{\rhx}{\varrho}
\newcommand{\si}{\sigma}
\newcommand{\six}{\varsigma}
\newcommand{\yp}{\upsilon}
\newcommand{\om}{\omega}
\newcommand{\phx}{\varphi}
\newcommand{\Ga}{\Gamma}
\newcommand{\De}{\Delta}
\newcommand{\Th}{\Theta}
\newcommand{\La}{\Lambda}
\newcommand{\Si}{\Sigma}
\newcommand{\Yp}{\Upsilon}
\newcommand{\Om}{\Omega}

%%%%%%%%%%%%%%%%%%%%%%%%%%%%%%%%%%%%%%%%%%%%%%%%%%%%%%%%%%%%%%%%%%%%%%%%%
%                         MATH MODE COMMAND                             %
%%%%%%%%%%%%%%%%%%%%%%%%%%%%%%%%%%%%%%%%%%%%%%%%%%%%%%%%%%%%%%%%%%%%%%%%%
\renewcommand{\L}{\cal{L}}
\newcommand{\M}{\cal{M}}
\newcommand{\be}{\begin{eqnarray}}
\newcommand{\ee}{\end{eqnarray}}
\newcommand{\jt}{\tilde{J}}
\newcommand{\Ra}{\Rightarrow}
\newcommand{\lra}{\longrightarrow}
\newcommand{\ti}{\tilde}
\newcommand{\pj}{\prod J}
\newcommand{\pjt}{\prod\tilde{J}}
\newcommand{\prb}{\prod b}
\newcommand{\prc}{\prod c}
\newcommand{\bft}{|\tilde{\phi}>}
\newcommand{\bfj}{|\phi>}
\newcommand{\lan}{\langle}
\newcommand{\ran}{\rangle}
\newcommand{\bz}{\bar{z}}
\newcommand{\bJ}{\bar{J}}
\newcommand{\vacr}{|0\rangle}
\newcommand{\vacl}{\langle 0|}
\newcommand{\IFF}{\Longleftrightarrow}
\newcommand{\phr}{|phys\ran}
\newcommand{\phl}{\lan phys|}
\newcommand{\nonu}{\nonumber\\}
\newcommand{\tg}{\tilde{g}}
\newcommand{\tM}{\ti{M}}
\newcommand{\hd}{\hat{d}}
\newcommand{\hL}{\hat{L}}
\newcommand{\sir}{\si^\rho}
\newcommand{\mf}{\mathfrak}
\newcommand{\mfg}{{\mathfrak{g}}}
\newcommand{\mfbg}{{\bar{\mathfrak{g}}}}
\newcommand{\mfk}{{\mathfrak{k}}}
\newcommand{\mfc}{{\mathfrak{c}}}
\newcommand{\mfbc}{{\bar{\mathfrak{c}}}}
\newcommand{\mfbk}{{\bar{\mathfrak{k}}}}
\newcommand{\mfn}{{\mathfrak{n}}}
\newcommand{\mfbn}{{\bar{\mathfrak{n}}}}
\newcommand{\mfh}{{\mathfrak{h}}}
\newcommand{\mfbh}{{\bar{\mathfrak{h}}}}
\newcommand{\mfp}{{\mathfrak{p}}}
\newcommand{\mfbp}{{\bar{\mathfrak{p}}}}
\newcommand{\mfU}{{\mf U}}
\newcommand{\Lg}{\L^{(\mfg})}
\newcommand{\Lgk}{\L^{(\mfg,\mfk)}}
\newcommand{\Lgkp}{\L^{(\mfg,\mfk')}}
\newcommand{\Lk}{\L^{(\mfk)}}
\newcommand{\Lkp}{\L^{(\mfk')}}
\newcommand{\Lc}{\L^{(\mfc)}}
\newcommand{\Lgg}{\L^{(\mfg,\mfg)}(\la)}
\newcommand{\Lgp}{\L^{(\mfg,\mfg')}(\la)}
\newcommand{\Mgp}{\M^{(\mfg,\mfg')}(\la)}
\newcommand{\Mgk}{\M^{(\mfg,\mfk)}(\la)}
\newcommand{\Hgk}{H^{(\mfg,\mfk)}}
\newcommand{\Mgkp}{\M^{(\mfg,\mfk^\prime)}}
\newcommand{\Hgp}{H^{(\mfg,\mfg')}}
\newcommand{\vo}{v_{0\la}}
\newcommand{\Ml}{\M(\la)}
\newcommand{\Ll}{\L(\la)}
\newcommand{\RR}{\mathbb{R}}
%%%%%%%%%%%%%%%%%%%%%%%%%%%%%%%%%%%%%%%%%%%%%%%%%%%%%%%%%%%%%%%%%%%%%%%%%
%                         MISCELLANEOUS                                 %
%%%%%%%%%%%%%%%%%%%%%%%%%%%%%%%%%%%%%%%%%%%%%%%%%%%%%%%%%%%%%%%%%%%%%%%%%

\newcommand{\e}[1]{\label{e:#1}\end{eqnarray}}
 \renewcommand{\r}[1]{(\ref{e:#1})}
 \newcommand{\ind}{\indent}
\newcommand{\np}{\newpage}
\newcommand{\hs}{\hspace*}
\newcommand{\vs}{\vspace*}
\newcommand{\nl}{\newline}
\newcommand{\bqu}{\begin{quotation}}
\newcommand{\equ}{\end{quotation}}
\newcommand{\bit}{\begin{itemize}}
\newcommand{\eit}{\end{itemize}}
\newcommand{\ben}{\begin{enumerate}}
\newcommand{\een}{\end{enumerate}}
\newcommand{\ul}{\underline}
\newcommand{\nn}{\nonumber}
\newcommand{\lef}{\left}
\newcommand{\rig}{\right}
\newcommand{\fra}{\twelvefrakh}
\newcommand{\Bb}{\twelvemsb}
\newcommand{\bT}{\bar{T}(\bz)}
\newcommand{\dagg}{^{\dagger}}
\newcommand{\qd}{\dot{q}}
\newcommand{\cP}{{\cal P}}
\newcommand{\hg}{\hat{g}}
\newcommand{\hh}{\hat{h}}
\newcommand{\hpg}{\hat{g}^\prime}
\newcommand{\htg}{\tilde{\hat{g}}^\prime}
\newcommand{\pri}{\prime}
\newcommand{\bis}{{\prime\prime}}
\newcommand{\lap}{\la^\prime}
\newcommand{\rhop}{\rho^\prime}
\newcommand{\Dgp}{\Delta_{g^\prime}^+}
\newcommand{\Dg}{\Delta_g^+}
\newcommand{\Pro}{\prod_{n=1}^\infty (1-q^n)}
\newcommand{\Pg}{P^+_{\hg}}
\newcommand{\Pgp}{P^+_{\hg\pri}}
\newcommand{\hmu}{\hat{\mu}}
\newcommand{\hnu}{\hat{\nu}}
\newcommand{\hrho}{\hat{\rho}}
\newcommand{\gp}{g^\prime}
\newcommand{\pp}{\prime\prime}
\newcommand{\CM}{\hat{C}(g',M')}
\newcommand{\CI}{\hat{C}(g',M^{\prime (1)})}
\newcommand{\CL}{\hat{C}(g',L')}
\newcommand{\HL}{\hat{H}^p (g',L')}
\newcommand{\HMI}{\hat{H}^{p+1}(g',M^{\prime (1)})}
\newcommand{\da}{\dagger}

\newcommand{\wro}{w^\rho}
\renewcommand{\Box}{\rule{2mm}{2mm}}
\begin{flushright}
June 23, 1998\\
HKS-NT-FR-98/3-SE\\
CERN-TH/98-173
\end{flushright}
\vs{10mm}

\begin{center}

{\Large{Unitarity of Strings and Non-compact Hermitian Symmetric Spaces}}
\\
\vspace{10 mm}
{\large{Stephen Hwang}\footnote{email: stephen.hwang@hks.se and
stephen.hwang@cern.ch}}
\vspace{4mm}\\
Department of Engineering Sciences, Physics and Mathematics,\\ Karlstad
University, S-651 88 Karlstad, Sweden\\and\\ Theory division, C.E.R.N.\\CH-1211
Geneva
23, Switzerland

\vs{15mm}

{\bf Abstract }  \end{center}
If $G$ is a simple non-compact Lie Group, with $K$ its maximal compact
subgroup, such
that $K$ contains a one-dimensional center $C$, then the coset space $G/K$ is
 an Hermitian symmetric non-compact space. $SL(2,\Bbb{R})/U(1)$ is the
simplest example of such a space. It is only when $G/K$ is an Hermitian
symmetric space that there exists unitary discrete representations of $G$. We
will
here study string theories defined as $G/K'$, $K'=K/C$, WZNW models.
We will establish unitarity for such string theories for certain discrete
representations. This proof
generalizes earlier results on $SL(2,\Bbb{R})$, which is the simplest example
of this class of theories. We will also prove unitarity of
$G/K$ conformal field theories generalizing results for $SL(2,\Bbb{R})/U(1)$.
We will
show that the physical space of states lie in a subspace of the
$G/K$ state space.
\vs{1cm}\\
{\em Accepted for publication in Physics Letters B}

\np
\setcounter{page}{1}
Although string theories have been extensively studied for several decades they 
still provide many surprises. String dualities are one of the more recent 
examples. In this paper we will study a class of string theories which provides 
one of the counterexamples to the successes of string theories. For string
theories propagating on non-compact curved backgrounds little is known although
some amount of effort has been put into investigations. The simplest example 
is the so-called $SL(2,\Bbb R)$ (or $SU(1,1)$ or $SO(2,1)$) string theory. This
theory is 
a $SL(2,\Bbb R)$ WZNW model with conformal anomaly $c=26$. For this case it was 
shown \cite{Hwa} that 
a selection of discrete representations similar to that of integrable 
representations for compact affine Lie algebras, gave unitarity in the physical 
subspace. In \cite{HHRS} a proposal of a modular invariant partition function 
was given. This proposal was further discussed in \cite{HR}. Still its status 
is an open question as it contains divergencies that seem to resist 
regularization. The question of consistency of the selection of representations
in an interacting theory
is furthermore an unresolved issue. A discussion has been given in
\cite{HR}. 
A proposal for another solution was given in \cite{Bars}.

Recent developments have renewed the interest in string theories on non-compact
curved manifolds. One of these is due to
Maldacena \cite{Mal}, who suggested that string theories on a $d+1$ dimensional
Anti 
de Sitter ($AdS$) space times some compact manifold is related to the large N 
limit of conformally invariant SU(N) gauge theories. Another development is due
to Strominger 
\cite{Str}, who made a
simple and elegant derivation of the Beckenstein-Hawking entropy formula for
three dimensional black holes, whose near-horizon symmetry is $AdS_3$. This
derivation uses the fact \cite{BH} that any consistent quantum gravity theory 
on $AdS_3$ is a two-dimensional conformal field theory (CFT). As $AdS_3$ is the 
group manifold of $SL(2,\Bbb R)$ it is suggestive that the relevant CFT giving
the 
correct number of states needed for the derivation in \cite{Str} is related to 
the $SL(2,\Bbb R)$ CFT. A derivation using this CFT was given by Carlip
\cite{Car}. 
A proposal for the CFT that describes $AdS_3$ has been given in \cite{BBG} and
involves a $SL(2,\Bbb R)/U(1)$ CFT.

In this work we will not attempt to address the unresolved issues of the 
$SL(2,R)$ string theory and related CFT's. Instead, we will focus on 
generalizations of these theories, with the hope that once the $SL(2,\Bbb R)$
model 
is fully understood the implications on the generalized theories considered 
here will be straightforward. There are two natural generalizations to the
$SL(2,\Bbb R)$ 
theory with one "time coordinate". One general class of theories
of which $SL(2,\Bbb R)$ is the simplest
example of 
is AdS spaces of arbitrary dimension with one "time-like" direction \cite{FL},\cite{BN}.
Such theories are of course highly interesting due to Maldacena's results. They
can be realized as the coset spaces $SO(d-1,2)/SO(d-1,1)$. 

Another class of
generalizations with one "time coordinate" is the 
following. Let $G$ be a non-compact Lie group and $K$ its maximal compact
subgroup. 
Require that $K$ contains a one-dimensional center $C$. The space
$G/K'$, $K'=K/C$, has one 
"time-like" (compact) direction and the simplest example is $SL(2,\Bbb R)$. The 
spaces $G/K$ are in these cases {\it Hermitian symmetric spaces}. A
classification of 
these spaces may be found in \cite{Hel}. There are four large classes: 
$SU(p,q)/S(U(p)\times U(q))$, $SO(p,2)/SO(p)\times SO(2)$, $SO^\ast(2n)/U(n)$ 
and $Sp(n,\Bbb R)/U(n)$. In 
addition there is one where $G$ is a certain real form of $E_6^{\Bbb C}$ and
another 
where it is a real form of $E_7^{\Bbb C}$. The dimension of the spaces are
$d=2pq,\ 2p,\
n(n-1),\ n(n+1),\ 32$ and $54$, respectively. Interestingly enough, it is
precisely when 
$G/K$, where $K$ refers to the maximal compact subgroup of $G$, is a 
Hermitian symmetric space that $G$ 
admits unitary highest weight modules \cite{HS1},\cite{HS2} (this is easily
understood from the fact that only in these cases can we choose the Cartan
subgroup to lie completely in $K$). These unitary representations have been 
completely classified in \cite{Ja}, \cite{EHW}. For any other 
non-compact group the unitary representations are necessarily continous. 
Comparatively
little is known about them in the general case. This is one of the reasons that
we have chosen
to study the second class of string theories. Another reason is the possibility
to generalize the $SL(2,\Bbb R)/U(1)$ construction, which is interesting due to
its interpretation as a two-dimensional black hole \cite{Wit}. The two above
classes, are the
only ones for $G$ simple which have one "time-coordinate" (see
\cite{GQ}, where also a number of
possibilities for $G$ being a product of simple factors are given).

Our main result, given in the Theorem, will be to establish unitarity for
string theories defined as $G/K'$ WZNW models times
some arbitrary unitary CFT
for certain highest
weight representations of $\mfg$, the affine Lie algebra
corresponding to $G$. These representations have highest weights $\la$ that are
antidominant. In the course of proving unitarity we will generalize
the result of Dixon, Lykken and Peskin \cite{DLP} on the unitarity of
$SL(2,\Bbb
R)/U(1)$ for discrete representations to $G/K$ for arbitrary Hermitian
symmetric pairs
$(G,K)$\footnote{In the string context $G/K'$ is not completely
arbitrary as the total conformal
anomaly satisfies $c=26$. This requires that $\mbox{dim}(G/K')\leq 26$. For the
unitarity of the coset $G/K$ there will be no such restriction.}. This
result is given in the Proposition. We will in fact prove that
the genuine physical state space is a subspace of the space defined by $G/K$. 

The techniques used to prove the Theorem are not the same as in the original
proof \cite{Hwa}. The reason is that the latter proof relied on the use of a
continous
basis, which is both unnatural and we do not know properties of its
generalization to the present case. Instead more relevant references to the
present case are \cite{BaH} and \cite{Hwa2} (relying to some extent on
techniques in \cite{GT}).
\vs{5mm}

We consider
$G$ to be 
a simple finite dimensional Lie group over $\Bbb R$ of
non-compact type and
$(G,K)$ a Hermitian symmetric pair.
We denote by $\mfbg_0$ and
$\mfbk_0$ the Lie
algebra of $G$ and $K$, respectively and
let $\mfbg_0=\mfbk_0+\mfbp_0$ be a Cartan decomposition i.e. a
decomposition into 
compact and non-compact pieces. Let
$\mfbh_0\subseteq\mfbk_0$ be a Cartan subalgebra (CSA). Then $\mfbh_0$
is also a CSA of $\mfbg_0$. Furthermore, $\mfbk_0=\mfbk_0'+\mfbc_0$, where 
$\mfbk'_0$ is a semi-simple algebra and
$\mfbc_0\in\mfbh_0$ is a one-dimensional center of $\mfbk_0$. Let
$\mfbg,\mfbk$ be the complexifications of $\mfbg_0,\mfbk_0$ etc. We denote
by $\bar{\De}$ the set
of roots of $\mfbg$ relative to $\mfbh$ and
$\bar{\De}=\bar{\De}_{\mfk}\cup\bar{\De}_{\mfp}$ is the
decomposition of
$\bar{\De}$ into compact roots
$\bar{\De}_{\mfk}$ and non-compact roots $\bar{\De}_{\mfp}$. We fix an
ordering of $\bar{\De}$ and
$\bar{\De}^+$ denotes the set of positive roots and
$\mfbg=\mfbn^-+\mfbh+\mfbn^+$ is a
triangular decomposition w.r.t. this ordering. Define
$\bar{\De}_{\mfp}^+\subset\bar{\De}^+$ and
$\bar{\De}_{\mfk}^+\subset\bar{\De}^+$ to be the corresponding positive
roots for $\mfbp$ and
$\mfbk$ and $\bar{\rho}=\sum_{\bar{\al}\in\bar{\De}^+}\bar{\al}$. Let
$\bar{\De}^s=\{\bar{\al}_1,\ldots,\bar{\al}_r\}\subset\bar{\De}^+$ be the
set of simple roots. It is always possible to 
choose an ordering of $\bar{\De}$ such
that $\bar{\al}_i>\bar{\al}_j$ for all $\bar{\al}_i\in\bar{\De}_{\mfp}$ and all
$\bar{\al}_j\in\bar{\De}_{\mfk}$. This is accomplished by taking the first
entry 
of $\bar{ \al}$ to be the eigenvalue with respect to $h\in\mfbc$. Since $\mfbc$ 
is in the center of $\mfbk$, roots of $\mfbk$ will have a zero as first entry. 
We have
$\bar{\al}_i+\bar{\al}_j\not\in\bar{\De}$ if
$\bar{\al}_i,\bar{\al}_j\in\bar{\De}^+_\mfp$. Furthermore, the highest root
$\bar{\th}$ of
$\bar{\De}$ is in $\bar{\De}_{\mfp}$ and $\bar{\De}^s\cap\bar\De_{\mfp}^+$
contains a single root. We define Hermite conjugation $x^\dagger$ of $x\in 
\mfg$ such that 
elements in $\mfbg_0$ are Hermitian. Then $h^\dagger=h$, $(e^{
\al})^\dagger=\pm e^{-\al}$, where the plus sign occurs when
$\bar{\al}\in\bar{\De}_{\mfk}$.

We extend the finite dimensional algebras $\mfbg_0$ and $\mfbg$ to the 
corresponding affine Lie algebras $\mfg_0$ and $\mfg$. The Cartan decomposition 
of $\mfbg$ extends naturally to a decomposition $\mfg=\mfk+\mfp$, where 
generators of $\mfg$ satisfy $h^\dagger=h$ for $h\in \mfh$, 
$(e^\al)^\dagger=\pm e^{-\al}$, where the plus sign occurs when 
$\al\in\De_\mfk$. $\mfk$ admits a further 
decomposition $\mfk=\mfk'+\mfc$ induced by the corresponding decomposition 
of the finite dimensional algebra. We denote by $\De$ the roots of $\mfg$ etc. 
for $\De_\mfk,\De_{\mfc}$. $\De^s$ and $\De^+$ denote simple and positive
roots, respectively. Denote by $\al_\mfc$ the roots 
$\al\in\De_\mfc^+$.

The string theory that we will consider here has a state space which is the
irreducible space $\Lgkp(\la) \times \M'$ of
$\Mgkp(\la)\times \M'$, where $\M'$ is the state space of an arbitrary unitary 
conformal field theory and $\Mgkp$ is formed by taking the highest weight 
$\mfg$ Verma module 
$\M^{(\mfg)}(\la)$ of 
highest weight $\la$ and requiring that the vectors $|v\ran \in \Mgkp(\la)$ are 
highest weight vectors with respect to $\mfk'$ i.e. $e^\al|v\ran=0$ for all
roots 
$\al\in\De_{\mfk'}^+$. In the simplest case,
$\mfg=\hat{sl}(2,\RR)$, $\M(\la)=\Mgkp(\la)$. The Virasoro generators $L_n$ 
relevant for our construction have the form
$L_n=L_n^{(\mfg,\mfk')} 
+L'_n$. Here
$L_n^{(\mfg,\mfk')}=L_n^{(\mfg)}-L_n^{(\mfk')}$ is the 
difference between the Virasoro generators formed from generators of $\mfg$ and 
$\mfk'$, respectively, in the usual way. We denote by
$L_n^{(\mfg,\mfk)}=L_n^{(\mfg,\mfk')}-L_n^{(\mfc)}$, where $L_n^{(\mfc)}$ is
the
time-like $\hat{u}(1)$ 
Virasoro generator, where time-like refers to the sign in the commutation
relations 
in the free field limit $k\rightarrow -\infty$. $k$ is here the central element
of $\mfg$. 
The central charge of the 
Virasoro algebra generated by $L_n$ is assumed to be 26. It is of the form
$c=c_\mfg-c_{\mfk'}+c'$, where
$c_\mfg={kd_\mfbg\over k+c_\mfbg}$, $c_{\mfk'}={kd_{\mfbk'}\over
k+c_{\mfbk'}}$ for $\mfk'$ simple and $c'$ is the central charge of the CFT defined on
$\M'$. Here $d_\mfbg$ is the 
dimension of $\mfbg$ and $c_\mfbg$ is the second 
Casimir of the adjoint representation of $\mfbg$ etc. for $\mfk'$. 
Reparametrization invariance of string theories requires that physical states 
$|\phi\ran$ should satisfy the Virasoro conditions
\be
L_n|\phi\ran=0,\hs{5mm}n>0,\e{1}
\be
(L_0-1)|\phi\ran=0.\hs{5mm}\e{2}
The result we will establish is the following.\vs{5mm}\\
%%%%%%%%%%%%%%%%%%%%%%%%%%
{\it Theorem.}  The solution of the eqs. \r{1} and \r{2} in $\Lgkp(\la)\times
M'$  with
$\la$ being antidominant are of the 
form
\be
|\phi\ran=|t\ran +|n\ran,\e{3}
where $|t\ran$ satisfies eqs. \r{1} and \r{2} and is a highest weight vector
w.r.t. $\mfk$ and $|n\ran$ 
is a physical null state, i.e. a state which decouples 
from all physical states. Furthermore, the states $|\phi\ran$ form a unitary
state
space.\vs{2mm}\\
%%%%%%%%%%%%%%%%%%%%%%%%%%%

Here a weight $\la$ is defined to be antidominant if $(2\la+\rho)\cdot\al\leq
0$ for 
all simple roots of $\mfg$, where $\rho$ is defined by $\rho\cdot\al=\al^2$. 
Antidominant weights require $k+c_\mfbg<0$. Note 
that although $|t\ran\in\Lgk(\la)\times
\M'$, the precise form of $\M'$ will be unimportant and we will henceforth
surpress its presence. The 
proof of the theorem will be divided into several steps. Our first step is to 
establish the following.\vs{5mm}\\
%%%%%%%%%%%%%%%%%%%%%%%%%%%%%%%%%%%
{\it Lemma 1.} A basis of states of $\Lgkp$ 
with $\la$ 
antidominant is
\be
|\psi\ran=L_{-n_1}\ldots L_{-n_k}
e^{-\al_1}\ldots e^{-\al_{k'}}|l, m\ran,\e{4}
where $n_1, \ldots n_k>0$, $\al_1,\ldots,\al_{k'}\in\De^+_\mfc$, $n_i\leq n_j$
and $\al_i\leq\al_j$ for $i<j$, and $|l, m\ran$
is highest 
weight w.r.t. $Vir(L)$, the Virasoro algebra generated by $L_n$, and $\mfc$
with weights $l$ and $m$, 
respectively.\vs{5mm}\\
{\it Proof.} Let $\cal{V}$ denote the largest irreducible subspace of $\Lgkp$
that has a basis of the following form
\be
|\psi\ran=L_{-n_1}^{(\mfg,\mfk)}\ldots L_{-n_k}^{(\mfg,\mfk)}
e^{-\al_1}\ldots e^{-\al_{k'}}|l^{(\mfg,\mfk)}\ran\otimes|m\ran,\e{5}
where $L_n^{(\mfg,\mfk)}=L_n-L_n^{(\mfc)}$ and
$|l^{(\mfg,\mfk)}\ran\otimes|m\ran$ is a highest
weight state w.r.t. $Vir(L^{(\mfg,\mfk)})+\mfc$ with weights
$l^{(\mfg,\mfk)},m$.
This space certainly exists. As $\cal{V}$ is irreducible the matrix 
$H(l^{(\mfg,\mfk)},m)$ of inner products restricted to vectors in $\cal{V}$ has
a 
non-vanishing determinant. Hence, we may form the orthogonal complement
$\cal{V}'$ to 
$\cal{V}$ with respect to the inner product. Assume that $\cal{V}'$ is
non-trivial and let 
$|v\ran\in \cal{V}'$ be a vector that has the 
smallest $L_0$ eigenvalue in $\cal{V}'$. Such a vector certainly
exists if
$\cal{V}'$
is 
non-trivial. Consider $|v_1\ran= x|v\ran$, where $x=L_n^{(\mfg,\mfk)}$ for some
value $n>0$ or $x=e^{\al_\mfc}$. Then $|v_1\ran\in \cal{V}$. If $|v_1\ran$ is 
zero for all
$n$ and all $\al_\mfc$, then $|v\ran\in\cal{V} $, and $\cal{V}'$ is 
trivial. Denote by $|v_2\ran$ a vector in $\cal{V}$ such that $\lan
v_2|v_1\ran\neq 
0$. $|v_2\ran$ exists as $\cal{V}$ is irreducible. We have $0\neq \lan 
(xv)|v_2\ran=\lan v|x^\dagger v_2\ran$. We will now prove that either
$x^\dagger|v_2\ran\in \cal{V}$ or it is zero. In both cases the r.h.s. is zero
(in the
former 
case from the orthogonality between $\cal{V}$ and $\cal{V}'$). Thus, we will
have a 
contradiction and $\cal{V}'$ is trivial.

Consider $|w\ran=x^\dagger|v_2\ran$ for any $|v_2\ran\in \cal{V}$. 
$|w\ran\in
\cal{V}$ except in the cases where $|w\ran$ is outside the irreducible
submodule. As 
the $\mfc$ Verma module is irreducible, this can only happen if there are
singular 
vectors in the $Vir(L^{(\mfg,\mfk)})$ module. The singular vectors may be
determined by 
reading off the zeros of
the Kac determinant corresponding to these modules. We have $c=25$ and for this
case the Kac determinant has zeros for 
$l^{(\mfg,\mfk)}=1-{(p+q)^2\over 4}$, where $p,q>0$ and integers. For
antidominant weights we have $l^{(\mfg,\mfk)}\geq 0$ with equality only for
$k\rightarrow -\infty$. This may be shown by carefully checking the explicit form of
$L_0^{(\mfg,\mfk)}$ (the first property, $l^{(\mfg,\mfk)}\geq 0$, follows, however, 
immediately from the
Proposition below). For finite values of $k$  the
Virasoro Verma module is irreducible and, hence, $x^\dagger|v_2\ran\in
\cal{V}$. For $k\rightarrow - \infty$ we have a null-vector for
$p=q=1$ and $l^{(\mfg,\mfk)} =0$. This
null-vector
is $|nv\ran=L^{(\mfg,\mfk)}_{-1}|l^{(\mfg,\mfk)}=0\ran$. The $\mfg$ module
becomes a Fock module as $k\rightarrow - \infty$ and in
this case $|nv\ran$ is identically zero. Thus, we have either 
$x^\dagger|v_2\ran\in \cal{V}$ or it is zero. We 
have proved our claim. The lemma follows now by replacing 
$L_n^{(\mfg,\mfk)}$ by $L_n$ using 
$L_n^{(\mfg,\mfk)}=L_n-L_n^{(\mfc)}$
and
commuting all the $L_n$'s to the left. 
\hs{5mm}\Box\vs{5mm}\\
%%%%%%%%%%%%%%%%%%%%%%%%%%
We may by \r{4} uniquely decompose $|\psi\ran$ as
\be
|\psi\ran=|\phi\ran+|s\ran\e{6}
Here $|\phi\ran\in\cal{P}$ is a state of the form $|\phi\ran=\prod 
e^{-\al_\mfc}|l\ran\otimes|m\ran$ and $|s\ran\in\cal{S}$ is a so-called 
spurious state 
i.e. of the form $|s\ran=L_{-n}|\chi\ran$ for some $n>0$ and $|\chi\ran$. The
next lemma is due to Goddard and Thorn
\cite{GT}.\vs{5mm}\\
%%%%%%%%%%%%%%%%%%%%%%%%%%%%%%%
{\it Lemma 2.} For $|\psi\ran$ of the form eq. \r{6}, the equations 
\be L_n|\psi\ran =(L_0-1)|\psi\ran=0,\hs{5mm} n>0,\e{7}
are equivalent for $c=26$ to
\be
L_n|\phi\ran=(L_0-1)|\phi\ran=0,\hs{4mm} &n>0,\nonumber\\
L_n|s\ran=(L_0-1)|s\ran=0.\hs{4mm} &n>0,\vs{5mm}\e{8}
%%%%%%%%%%%%%%%%%%%%%%%%%%%%%%%%
{\it Lemma 3.} A basis of $\L^{(\mfg)}(\la)$, the irreducible module of
$\M^{(\mfg)}(\la)$, is for antidominant weights $\la$
\be
|v\ran=e^{-\al_1}\ldots e^{-\al_n}|\la_{\mfk}\ran, \hs{5mm}
\al_1,\ldots\al_n\in\De^s_{\mfk}\cup\De^+_\mfc.\e{9}
Here $|\la_{\mfk}\ran$ is a highest weight state w.r.t. $\mfk$ with weight 
$\la_{\mfk}$.\vs{5mm}\\
{\it Proof.} The Kac-Kazhdan determinant formula for states with weight
$\la-\eta$ in $\M^{(\mfg)}(\la)$ 
reads \cite{KK}
\be\mbox{det}H_\eta=C_1\prod_{\al\in\De^+}\prod_{n=1}^\infty[(2\la+\rho,\al)-
n(\al,\al)]^{P(\eta-n\al)},\e{10}
where $C_1$ is a non-zero constant. From this formula it is evident that for 
$\la$ being antidominant $\M^{(\mfg)}(\la)$ is irreducible i.e. 
$\M^{(\mfg)}(\la)$ and 
$\L^{(\mfg)}(\la)$ are isomorphic. Denote by $\cal{V}$ the largest irreducible
space 
that 
has \r{9} as a basis. As $\M^{(\mfg)}(\la)$ is irreducible it follows that if 
$|v\ran\in 
\cal{V}$ then $x^\dagger |v\ran\in \cal{V} $ for any $x\in \mfk$. By exactly the
same argument 
that lead us to conclude that \r{5} was a basis, 
we may conclude that the orthogonal complement $\cal{V}'$ to $\cal{V}$ is
trivial
and hence 
\r{9} is a basis.\hs{5mm}\Box\vs{5mm}\\
{\it Proposition.} The space $\L^{(\mfg,\mfk)}(\la)$ is unitary for $\la$ being 
antidominant.\vs{5mm}\\
{\it Proof.} We have by Lemma 3 that the Kac-Kazhdan determinant for 
$\M^{(\mfg)}(\la)$ can be written in the form
\be
\mbox{det}H=\mbox{det}H^{(\mfk)}\mbox{det}H^{(\mfg,\mfk)}
\e{11}
Here $H^{(\mfg,\mfk)}$ arises from states in $\L^{(\mfg,\mfk)}$ and 
$H^{(\mfk)}$ from all other states, which by Lemma 3 are non-highest weight 
states in $\mfk$ modules. As $\mbox{det}H$ is non-zero it follows that 
$\mbox{det}H^{(\mfg,\mfk)}$ is non-zero. This implies that the signature of 
$H^{(\mfg,\mfk)}$ remains unchanged for any antidominant value of $\la$. We 
consider the 
limit $k\rightarrow -\infty $, which is a possible limit for antidominant
weights. 
This limit is just the free 
field limit in which $\mfg$ reduces to a direct sum of Fock algebras. Then 
a basis of $\L^{(\mfg,\mfk)}$ are states of the form
\be
|v^{(\mfg,\mfk)}\ran=e^{\prime -\al_1}\ldots e^{\prime -\al_n}|\la\ran,
\hs{5mm}\al_i\leq\al_j\mbox{ for }i\leq j,\hs{5mm}
\al_1,\ldots,\al_n\in\De^+_\mfp,\e{12}
where $e^{\prime -\al_i}={1\over \sqrt{-k}}e^{-\al_i}$ and $|\la\ran$ is the 
highest weight state generating $\M^{(\mfg)}(\la)$. As $e^{\prime -\al_i}$ act
as free
field 
modes with a positive signature, it is  
evident that $H^{(\mfg,\mfk)}$ has a positive definite signature in this limit 
and hence for all antidominant weights.\hs{5mm}\Box\vs{5mm}\\
As remarked in the proof of Lemma 1, an immediate consequence of the 
Proposition is that $L_0^{(\mfg,\mfk)}$ has non-negative eigenvalues in $\Lgk$.
The Proposition gives a generalization of the unitarity theorem given in
\cite{DLP} for the case $SL(2,\Bbb R)/U(1)$. Notice that the Proposition is
true even for $c\neq 26$. \vs{5mm}\\
%%%%%%%%%%%%%%%%%%%%%%%%%%%%
{\it Proof of Theorem.} By Lemma 1 and Lemma 2 we have that physical states 
should satisfy 
eq. \r{8}. As any state in $\cal{S}$ decouples from all physical states 
it is sufficient to study states in $\cal{P}$ satisfying eq. \r{8}. By Lemma 1 
we can choose a basis of $\cal{P}$ consisting of states $|t\ran\in\cal{T}$,
which are highest weight states w.r.t. the algebra $Vir(L)+\mfc$,
and $|u\ran\in\cal{U}$ which 
are created from $|t\ran$ by acting with $e^{-\al_\mfc}$. Decomposing
$|\phi\ran\in 
\cal{P}$ as $|\phi\ran=|t\ran+|u\ran$ it follows that $L_n|\phi\ran=0$ reduces 
to the equation $L_n|u\ran=0$, which in turn reduces to $L_n^{(\mfc)}|u\ran=0$.
The $\mfc$ module is a Fock module and it is well-known that there are no 
non-trivial on-shell states, except for $m=0$ \footnote{This special case 
was overlooked in \cite{Hwa2}}.  For $m=0$ we have a
state $|u_1\ran=e^{-\al_\mfc}|l, m=0\ran$, where
$|l, m=0\ran\in\cal{T}$ and $[L_0,e^{-\al_\mfc}]=e^{-\al_\mfc}$,
which satisfies $L_n|u_1\ran=0$, $n>0$. The on-shell condition requires $l=0$ so that
$l^{(\mfg,\mfk)}=0$, which in turn requires $k\rightarrow -\infty$ 
(cf. the proof of Lemma 1). In this free
field limit it is easy to see and well-known that there are no on-shell states
in $\cal{U}$
that have $m=0$. Thus in the space $\cal{P}$, only states in 
$\cal{T}$ are physical. This is the first claim of the theorem. The second 
claim is that this subspace is unitary and it follows from the 
Proposition.\hs{5mm}\Box
\vs{5mm}\\
Note that $\cal{T}$ is a subspace of $\Lgk$ and, hence, the genuine physical
states are in $\Lgk$.

One relevant question is whether it is necessary to assume antidominant weights
for unitarity. If one considers the free field limit $k\rightarrow -\infty$,
then it is easily seen that a basis of $\cal{L}^{(\mfg,\mfk)}$ is given by
states
of the form $e^{\prime -\al_{1}}\ldots e^{\prime -\al_{n}}|\la\ran$, $\al_i\in\De^+_\mfp$.
Unitarity of this state space then requires $\la\cdot\al\leq 0$ for
$\al\in\De^+_\mfp$. Our conjecture is, therefore, that it is sufficient to
require $(2\la+\rho)\cdot\al\leq 0$, $\al\in\De^+_\mfp$ to ensure
unitarity.\vs{10mm}\\
{\bf Acknowledgement:} I would like to thank Steve Carlip and Genkai Zhang
for helpful discussions.\vs{1cm}


\newpage

\begin{thebibliography}{AA}
\bibitem{Hwa}S. Hwang, Nucl. Phys. {\bf B351} (1991) 425 
\bibitem{HHRS}M. 
Henningson, S. Hwang, P. Roberts and B. Sundborg, Phys. Lett. {\bf B267}
(1991) 350
\bibitem{HR}S. Hwang and P. Roberts, Proceedings of the 16'th Johns Hopkins' 
workshop, ed. by L. Brink and R. Marnelius, World Scientific 
1993 (hep-th/9211075)
\bibitem{Bars} I. Bars, Phys. Rev. {\bf D53} (1996) 3308
\bibitem{Mal} J. Maldacena, hep-th/9711200
\bibitem{Str} A. Strominger, hep-th/9712251
\bibitem{BH} J.D. Brown and M. Henneaux, Comm. Math. Phys. {\bf 104} (1986) 207
\bibitem{Car} S. Carlip, Phys. Rev. {\bf D51} (1995) 632
\bibitem{BBG} K. Behrndt, I. Brunner and I. Gaida, hep-th/9804159
\bibitem{FL} E.S. Fradkin and V.Ya. Linetsky, Phys. Lett. {\bf B261} (1991) 26 
\bibitem{BN} I. Bars, D. Nemenschansky, Nucl. Phys. {\bf B348} (1991) 89
\bibitem{Hel}S. Helgason, {\it Differential geometry, Lie groups, and
symmetric spaces},
 Academic Press, 1978
\bibitem{HS1}Harish-Chandra, 
Amer. J. Math. {\bf 77} (1955), 743
\bibitem{HS2}Harish-Chandra,
Amer. J. Math. {\bf 78} (1956), 1
\bibitem{Ja}H.P. Jakobsen,  J. Funct. Anal. {\bf 52} (1983) 385-412
\bibitem{EHW}T. Enright, R. Howe and N. Wallach, Representation theory of
reductive groups, Proceedings of
the university
of Utah Conference 1982, Birkh\"auser, Boston 1983, p 97-143
\bibitem{Wit} E. Witten, Phys. Rev. {\bf D44} (1991) 314
\bibitem{GQ} P. Ginsparg and F. Quevedo, Nucl. Phys. {\bf B385} (1992) 527
\bibitem{DLP}L.J. Dixon, J. Lykken and M.E. Peskin,  Nucl. Phys. {\bf B325}
(1989) 329
\bibitem{BaH}C. Bachas and S. Hwang, Phys. Lett. {\bf B247} (1990) 265
\bibitem{Hwa2} S. Hwang, Phys. Lett. {\bf B276} (1992) 451
\bibitem{GT}P. Goddard and C. Thorn, Phys. Lett. {\bf B40} (1972) 
235
\bibitem{KK} V.G. Kac and D.A. Kazhdan, Adv. in Math. {\bf 34} (1979) 97


\end{thebibliography}
\end{document}